\documentclass[%
 reprint,
 amsmath,amssymb,
 aps,
]{revtex4-2}

\usepackage{graphicx}% Include figure files
\usepackage{dcolumn}% Align table columns on decimal point
\usepackage{bm}% bold math
\usepackage[colorlinks=true,linkcolor=blue,citecolor=blue,urlcolor=blue]{hyperref}

\begin{document}

\preprint{APS/123-QED}

\title{Periodically poled lithium niobate with submicrometer periodicity using electric-field poling technique}% Force line breaks with \\

\author{Yuchen Zhang}%
\affiliation{The MOE Key Laboratory of Weak Light Nonlinear Photonics, School of Physics and TEDA Applied Physics Institute, Nankai University, Tianjin 300071, China}%

\author{Sanbing Li}%
\affiliation{The MOE Key Laboratory of Weak Light Nonlinear Photonics, School of Physics and TEDA Applied Physics Institute, Nankai University, Tianjin 300071, China}%

\author{Shengbo Song}%
\affiliation{The MOE Key Laboratory of Weak Light Nonlinear Photonics, School of Physics and TEDA Applied Physics Institute, Nankai University, Tianjin 300071, China}%

\author{Xitan Xu}%
\affiliation{The MOE Key Laboratory of Weak Light Nonlinear Photonics, School of Physics and TEDA Applied Physics Institute, Nankai University, Tianjin 300071, China}%

\author{Xudong Tong}%
\affiliation{The MOE Key Laboratory of Weak Light Nonlinear Photonics, School of Physics and TEDA Applied Physics Institute, Nankai University, Tianjin 300071, China}%

\author{Rui Yang}%
\affiliation{The MOE Key Laboratory of Weak Light Nonlinear Photonics, School of Physics and TEDA Applied Physics Institute, Nankai University, Tianjin 300071, China}%

\author{Meiyu Liu}%
\affiliation{The MOE Key Laboratory of Weak Light Nonlinear Photonics, School of Physics and TEDA Applied Physics Institute, Nankai University, Tianjin 300071, China}%

\author{Jiangdong Sun}%
\affiliation{The MOE Key Laboratory of Weak Light Nonlinear Photonics, School of Physics and TEDA Applied Physics Institute, Nankai University, Tianjin 300071, China}%

\author{Feng Gao}%
\affiliation{The MOE Key Laboratory of Weak Light Nonlinear Photonics, School of Physics and TEDA Applied Physics Institute, Nankai University, Tianjin 300071, China}%

\author{Fang Bo}%
\affiliation{The MOE Key Laboratory of Weak Light Nonlinear Photonics, School of Physics and TEDA Applied Physics Institute, Nankai University, Tianjin 300071, China}%

\author{Jingjun Xu}%
\affiliation{The MOE Key Laboratory of Weak Light Nonlinear Photonics, School of Physics and TEDA Applied Physics Institute, Nankai University, Tianjin 300071, China}%

\author{Guoquan Zhang}%
\email{zhanggq@nankai.edu.cn}
\affiliation{The MOE Key Laboratory of Weak Light Nonlinear Photonics, School of Physics and TEDA Applied Physics Institute, Nankai University, Tianjin 300071, China}%

\date{\today}% It is always \today, today,
             %  but any date may be explicitly specified

\begin{abstract}
A series of monocrystalline lithium niobate (LN) thin films on low-resistivity silicon (LNOS) substrates, with thicknesses ranging from micrometers down to tens of nanometers, were prepared through chemical-mechanical polishing of pristine crystal films. By employing the electric-field poling technique, high-quality periodically poled lithium niobate (PPLN) structures with submicrometer periodicity were fabricated in these LNOSs, and full-thickness polarization reversal beneath the poling electrodes with a well-defined depth profile was achieved. The coercive field was found to follow the Kay-Dunn scaling law, while the lateral domain broadening generally increased with increasing film thickness, and it remained approximately constant across different electrode linewidths at a given film thickness. PPLN structures with periods ranging from 6 $\mu$m down to 200 nm were demonstrated. These findings provide guidance for poling electrode design and lateral domain broadening pre-compensation, offering a practical route toward scalable domain engineering in LN thin films with promising potential for compact and highly integrated nonlinear photonic devices.
\end{abstract}

\maketitle

%\tableofcontents

\section{\label{sec:level1}Introduction}

Periodically poled lithium niobate (PPLN) is a fundamental structure in modern photonics and has enabled a wide range of applications, including nonlinear frequency conversion~\cite{wang2018ultrahigh,lin2019broadband,lu2020toward}, optical parametric oscillation~\cite{myers1995quasi,kartalouglu2003simultaneous}, and on-chip entangled photon pair generation~\cite{jin2014chip}. In addition to these optical functionalities, polarization reversal in LN can also create ferroelectric conductive domain walls (DWs), which serve as nanoscale conductive channels embedded in an insulating matrix and have been explored for nanoelectronic devices such as reconfigurable memristors~\cite{lu2019electrical,suna2023tuning}, nonvolatile transistors~\cite{chai2020nonvolatile}, and domain-wall-based $p$-$n$ junctions~\cite{qian2022domain,li2025nanoscale}. With the advent of monocrystalline LN films enabled by ion-slicing techniques~\cite{jia2021ion,zhu2021integrated}, PPLN has become an increasingly important functional element for compact integrated photonic and electronic devices, thereby placing growing demands on precise domain engineering across multiple length scales.

Several approaches have been developed for the fabrication of PPLN structures, including electric-field poling~\cite{shur2015micro,qian2022domain,zhang2022conductive}, direct laser writing~\cite{valdivia2005nanoscale,muir2008direct,shur2011situ,xu2022femtosecond}, electron- or ion-beam direct writing~\cite{shur2014domain,chezganov2017influence,kokhanchik2022morphology}, and field-enhanced scanning-probe-based domain inversion techniques~\cite{jiao2020improvement,ievlev2014intermittency,qian2022graphical,song2026jap}. Among these methods, electric-field poling remains the most mature and scalable approach for producing periodic domain structures over large areas, offering good controllability, reproducibility, and compatibility with practical device fabrication. Compared with other techniques, electric-field poling generally avoids the direct laser-induced thermal effects and potential beam-induced damage associated with laser and charged-particle-beam writing, respectively, while offering higher fabrication throughput than scanning-probe-based domain inversion. Nevertheless, electric-field poling of LN films still faces significant challenges. Polarization reversal requires an applied electric field exceeding the coercive field, which is typically around 21 kV/mm for congruent LN~\cite{lines2001principles}. During polarization switching, domain penetration through the film thickness is generally accompanied by lateral domain expansion, which can lead to pronounced domain broadening~\cite{zhao2019optical,lu2019periodically,stanicki2020surface,liu2024promoting,sabatti2025nanodomain}. Moreover, the inverted domains often exhibit irregular morphologies that are difficult to control~\cite{mackwitz2016periodic,nagy2019reducing,stanicki2020surface,wu2022broadband,su2024high,yang2024symmetric}. These challenges become even more severe in the submicrometer regime, where high-quality PPLN fabrication requires precise control over both vertical domain penetration and lateral domain expansion. Consequently, the fabrication of high-quality submicrometer-period PPLN structures remains technically demanding, limiting their further development for applications such as backward second-harmonic generation, mirrorless optical parametric oscillation, and narrowband quantum light sources.

\begin{figure*}[ht]
	\centering
	\includegraphics[width=15cm]{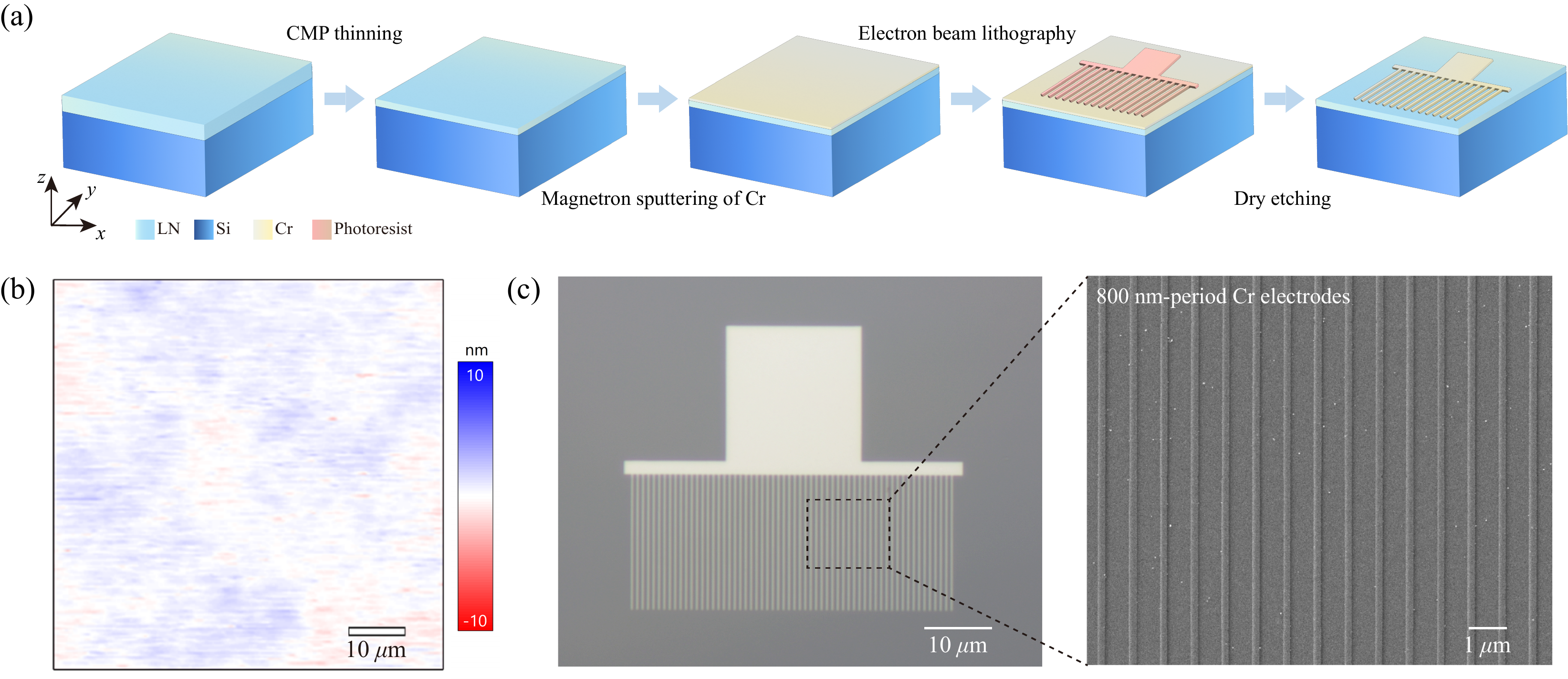}%
	\caption{\label{Fig1} Fabrication and characterization of LNOS and periodic poling electrodes. (a) Schematic diagram of the film structure and the electrode-patterning process. (b) The 2D surface topography of the CMP-milled LN films measured by AFM. (c) SEM image of the poling electrodes, with an adjacent enlarged view showing the Cr electrodes with a period of 800 nm.}
\end{figure*}

In the commonly used $z$-cut monocrystalline LN films investigated in this work, where the spontaneous polarization $\mathbf{P}_s$ is oriented perpendicular to the film plane, the fabrication of submicrometer-period PPLN involves a trade-off between lateral confinement and depthwise control of the inverted domains. In the absence of an electrode interlayer beneath the LN film, domain inversion by electric-field poling typically requires relatively high voltages, usually on the order of several hundred volts. Increasing the applied voltage can facilitate domain growth along the film thickness, but it may also aggravate lateral domain broadening. Conversely, reducing the poling voltage or shortening the voltage pulse duration can suppress domain broadening, yet often leads to surface-confined inversion or conical domain profiles~\cite{yu2024poling,yang2024symmetric}. For example, Yang $et$ $al$. successfully demonstrated PPLN structures with periods down to 370 nm, while the inverted domains exhibited a conical depth profile that could reduce the harmonic-generation efficiency relative to that predicted for an ideal domain structure~\cite{yang2024symmetric}. Field-enhanced scanning-probe-based methods can realize PPLN structures with submicrometer periodicity and provide better control over the shape of the domains; however, the resulting domains often fail to penetrate through the entire film thickness, limiting their suitability for applications requiring uniform full-thickness polarization reversal~\cite{lu2019electrical,slautin2020domain}. 

More recently, building on their earlier demonstration of 370-nm-period PPLN structures, Yang $et$ $al$.  demonstrated backward-phase-matched optical parametric oscillation and simultaneous Type-0 and Type-I parametric processes in integrated LN microresonators~\cite{Yang2026nano,Yang2026np}. These advances highlight the growing potential of PPLN structures with submicrometer periodicity for integrated nonlinear photonic applications and underscore the importance of precise domain engineering. Nevertheless, achieving submicrometer-period PPLN structures with both high lateral fidelity and uniform full-thickness polarization reversal remains a great challenge for LN thin films.

In this work, we demonstrate the fabrication of high-quality PPLN structures with submicrometer periodicity in a series of LN films on low-resistivity silicon (LNOS) substrates through electric-field poling technique. The thickness-dependent polarization switching and lateral domain broadening are systematically investigated to guide poling electrode design, and PPLN structures with periods ranging from 6 $\mu$m down to 200 nm are successfully fabricated. As a typical example without loss of generality, a 3-mm-long 800-nm period PPLN structure with a pre-designed duty cycle and full-thickness polarization reversal of well-defined domain profiles is demonstrated.  These results provide a practical route toward scalable submicrometer domain engineering in LN thin films, with promising potential for compact and highly integrated nonlinear photonic devices.

\begin{figure*}[htbp]
	\centering
	\includegraphics[width=15.5cm]{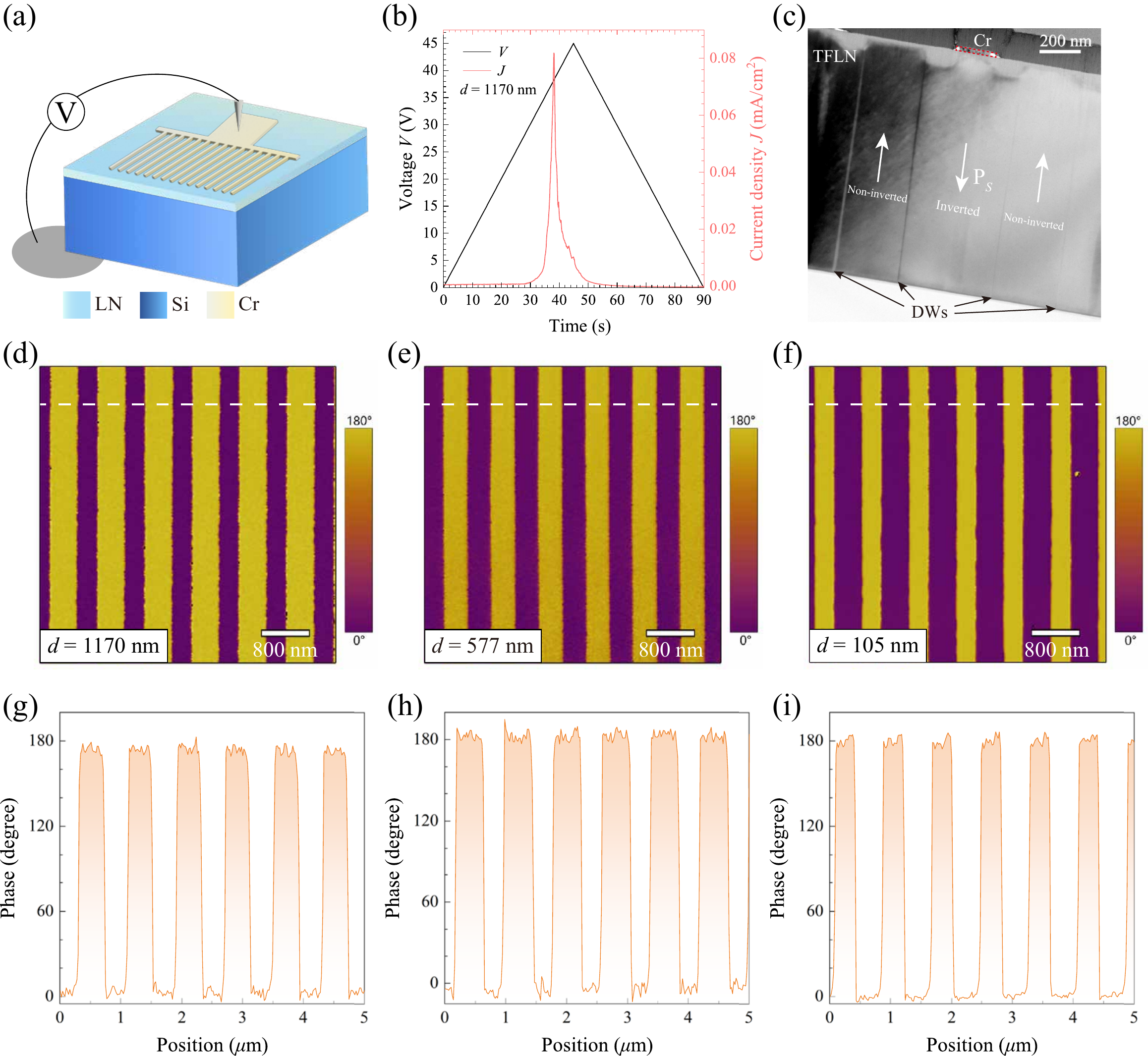}%
	\caption{\label{Fig2} Fabrication and characterization of PPLN structures. (a) Schematic diagram of the electric-field poling configuration. (b) Temporal evolution of the applied voltage $V$ and the resulting poling current density $J$ for the LNOS with a LN film thickness $d$ of 1170 nm. (c) Cross-sectional TEM image of the inverted domains in a 1170-nm-thick LN film. The white and black arrows indicate the directions of $\mathbf{P}_s$ and the locations of the DWs, respectively. The red dashed box corresponds to the location of the Cr electrode. (d-f) PFM phase images of 800-nm-period PPLN structures fabricated in LNOSs with LN film thicknesses $d$ of 1170, 577, and 105 nm, respectively. (g-i) Corresponding phase distribution profiles of the PPLN structures extracted along the white dashed lines in (d-f), respectively. The film thickness $d$ is indicated in the bottom-left corner of each panel in (d-f), and the yellow regions represent the inverted domains.}
\end{figure*}

\section{Results and Discussions}

The pristine structure consisted of a $z$-cut nominally pure congruent monocrystalline LN film with a thickness of 2 $\mu$m (the top surface was the +$z$ surface), and a 400-$\mu$m-thick low-resistivity silicon substrate (with a resistivity of 0.02 $\Omega$$\cdot$cm) that were bonded directly one another, as shown schematically in Fig.~\ref{Fig1}(a). The bonded wafer was commercially supplied by NANOLN and subsequently diced into pieces, each with an area of 10$\times$10 mm$^2$. Thin films with different thicknesses were prepared by utilizing the chemical-mechanical polishing (CMP) technique to mill the pristine films with different polishing time while keeping other parameters the same, such as the mechanical pressure on the pristine films, the concentration of the polishing solution (MasterMet, available from Buehler), and the rotation speed (30 rpm). Consequently, a series of films with thicknesses ranging from the micrometer scale down to the nanoscale were obtained. The film thickness $d$ was measured using a reflection-based film-thickness measurement system (Apris A4-SRM), and can be determined by nonlinear fitting of the measured reflection spectra, based on the interference of light reflected from the film interfaces. Prior to film characterization and PPLN fabrication, the films were cleaned in piranha solution (7:3 98\% H$_2$SO$_4$:30\% H$_2$O$_2$) at 75$^\circ$C for 2 minutes, followed by rinsing with distilled water and drying under a stream of nitrogen. The surface topography of the CMP-milled thin films was characterized in two dimensions (2D) by atomic force microscope (AFM; Asylum Research, Goleta, CA, USA) operated in contact mode, as shown in Fig.~\ref{Fig1}(b). The surface roughness was measured to be 0.5 nm. 

\begin{figure*}[htbp]
	\centering
	\includegraphics[width=12cm]{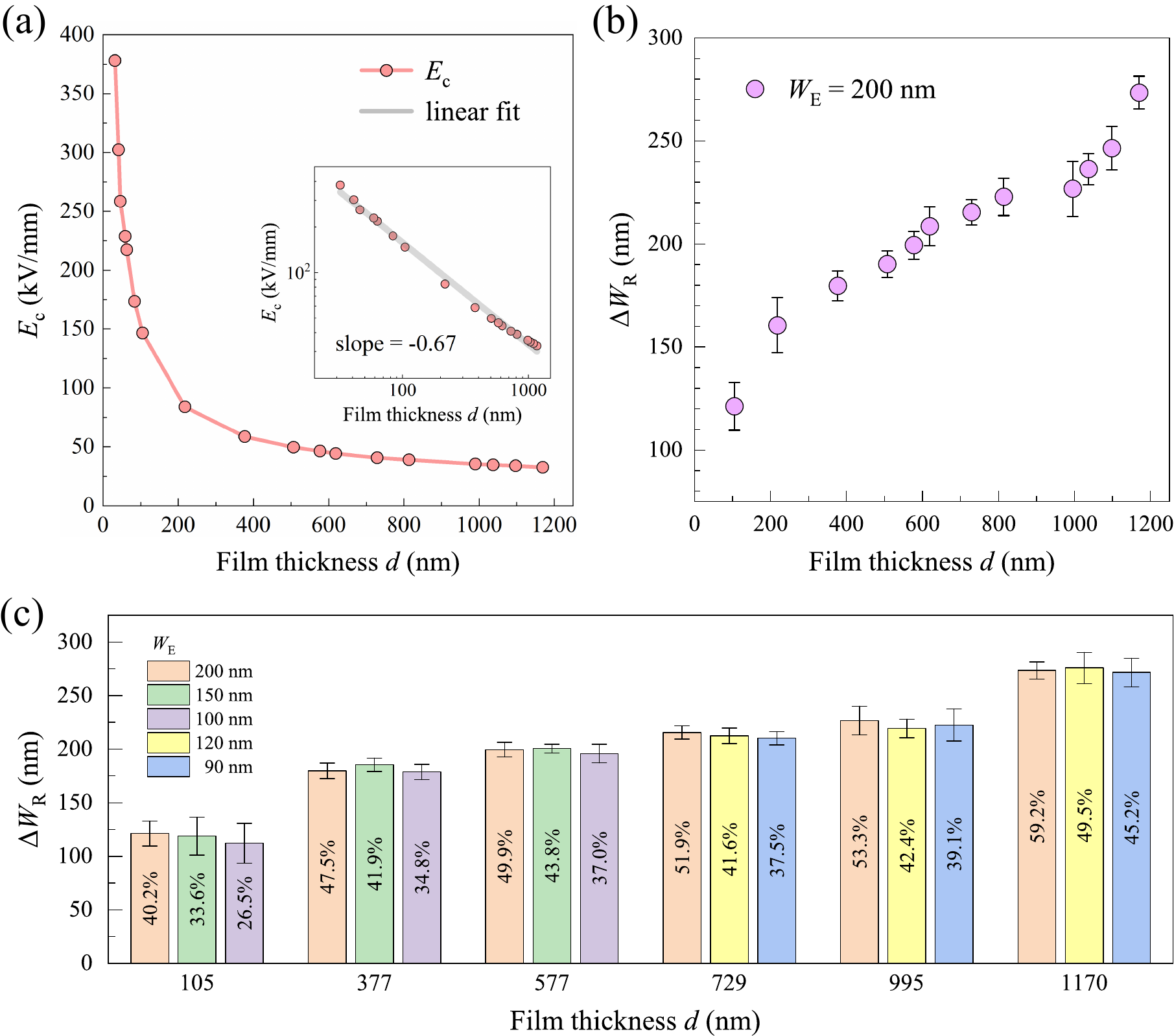}%
	\caption{\label{Fig3} Thickness-dependent polarization switching characteristics of the fabricated PPLN structures. (a) Dependence of the coercive field $E_{\rm{c}}$ on the film thickness $d$. The solid line in the inset represents a linear fit to the experimental data on a logarithmic scale. (b) Dependence of the inverted-domain broadening $\Delta W_{\rm{R}}$ on the film thickness $d$, with the poling electrode linewidth $W_{\rm{E}}$ fixed at 200 nm. (c) Inverted-domain broadening $\Delta W_{\rm R}$ for different film thicknesses and poling electrode linewidths. The corresponding duty cycles are indicated inside the bars. Error bars in (b) and (c) represent standard deviations. In all cases, the period of the PPLN structures is 800 nm.}
\end{figure*} 

Poling electrodes were patterned on the top $+z$ surface of LN films through a sequence of magnetron sputtering, electron-beam lithography, and dry etching, as described in Fig.~\ref{Fig1}(a). For the target PPLN structure with a period of 800 nm, the Cr poling electrodes consisted of 60 comb fingers, each with a thickness of 20 nm, a linewidth of 200 nm, and a length of 20 $\mu$m. A square contact pad with a side length of 20 $\mu$m was connected to the comb electrodes through a 50-$\mu$m-long and 2-$\mu$m-wide Cr stripe to provide electrical contact to the external circuit. Figure~\ref{Fig1}(c) shows a scanning electron microscope (SEM) image of the overall electrode structure, together with an enlarged view of the Cr electrodes with an 800-nm period.

The fabrication process of PPLN structures is schematically illustrated in Fig.~\ref{Fig2}(a). During the poling process, the applied voltage was supplied by an ammeter (Keithley 6517B), and the poling current was recorded simultaneously. The top Cr electrode was connected to the voltage source through a probe equipped with a flexible tungsten needle (GGB Industries, T-4-35), and the bottom low-resistivity silicon substrate was connected to the low terminal of the ammeter through conductive silver paste and a probe equipped with a hard tungsten needle (GGB Industries, ST-20-10). The low terminal served as the ground reference, and the current was defined as positive when flowing from the top electrode to the bottom substrate. The applied voltage was linearly increased to the target value at a rate of 1 V/s and then decreased to zero at the same rate. Unless otherwise specified, the thickness of the Cr poling electrodes and the electrical poling configuration were kept unchanged throughout this work, while the electrode linewidth and period were varied to obtain PPLN structures with different duty cycles and periods. Figure~\ref{Fig2}(b) shows the temporal evolution of the applied voltage and the resulting poling current density for the LNOS with an LN film thickness $d$ of 1170 nm. Similar poling current responses were also observed in LNOSs with other LN film thicknesses. Notably, the poling current gradually diminished as the voltage increased further, indicating that polarization reversal within the electrode-defined region had largely completed~\cite{ya2014micro}. To examine the depth profile of the inverted domains, cross-sectional transmission electron microscope (TEM) was performed, as DWs can be clearly resolved in TEM images~\cite{chai2020nonvolatile}. Figure~\ref{Fig2}(c) presents a cross-sectional TEM image of the 800-nm-period PPLN structures fabricated in a 1170-nm-thick LN film. As indicated by the black arrows, the DWs exhibit well-defined, nearly straight profiles extending through the entire film thickness, providing direct evidence of full-thickness polarization reversal beneath the poling electrodes.

Following polarization reversal, the Cr electrodes were removed using chrome etchant (Chrome Etchant Type 1020), and the resulting domain structures were characterized by piezoresponse force microscope (PFM). As shown in Figs.~\ref{Fig2}(d-f), PPLN structures with a period of 800 nm were successfully fabricated in LNOSs with LN film thicknesses $d$ of 1170, 577, and 105 nm, respectively, where the yellow regions correspond to the inverted domains. The phase distribution profiles along the white dashed lines in Figs.~\ref{Fig2}(d-f) are displayed in Figs.~\ref{Fig2}(g-i), respectively, further confirming the high structural quality and accurate domain configuration of the fabricated PPLN structures.

\begin{figure*}[htbp]
	\centering
	\includegraphics[width=17cm]{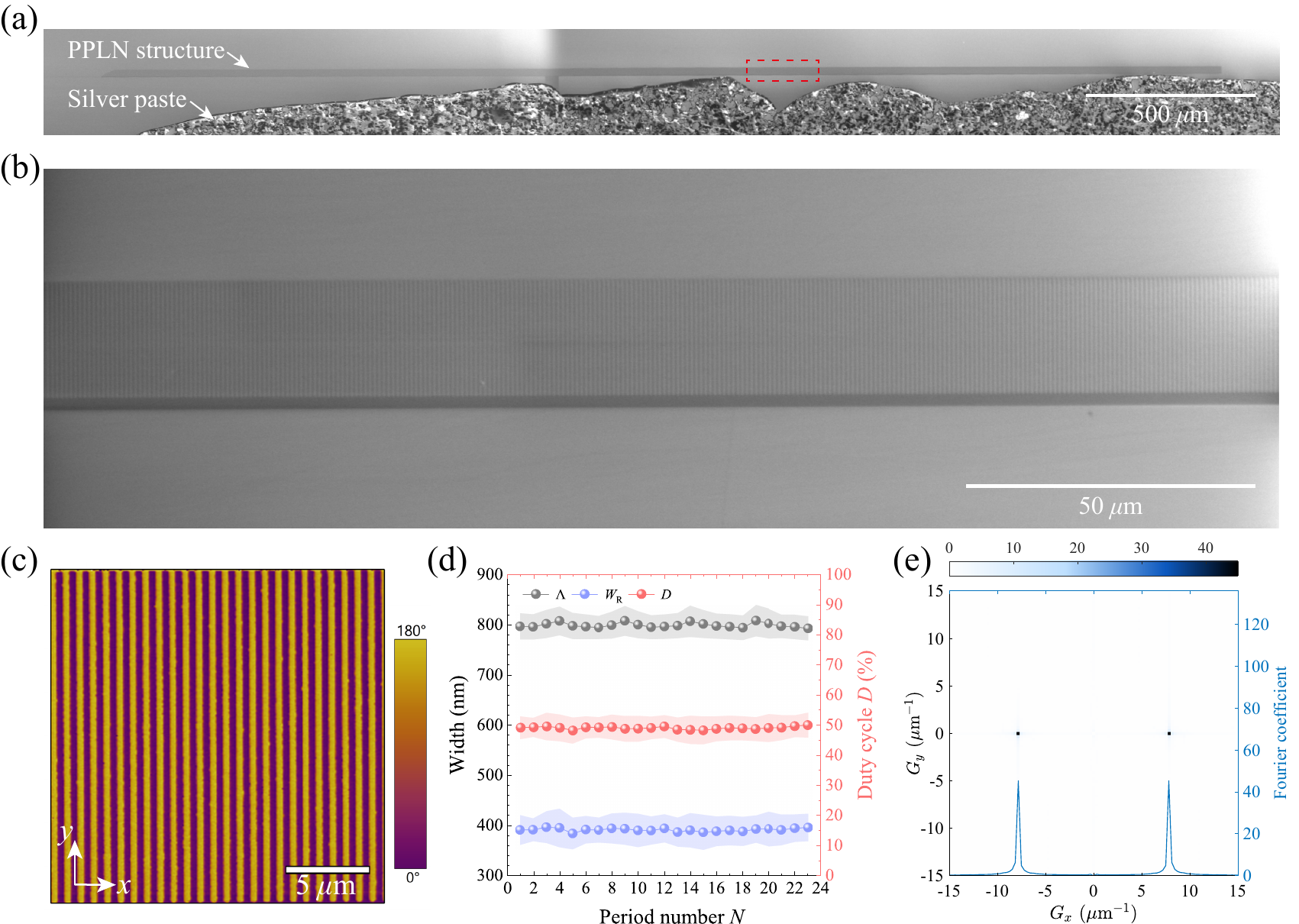}%
	\caption{\label{Fig4} Fabrication and characterization of millimeter-long PPLN structure with an 800-nm period in LNOS. (a) Overview SEM image of the 3-mm-long PPLN structure. Silver paste was applied to improve electrical conduction during SEM imaging of the PPLN structure. (b) Enlarged SEM image of the region marked by the red box in (a), showing the periodic domain pattern. (c) PFM phase image acquired in the middle of the PPLN structure over a 20 $\mu$m $\times$ 20 $\mu$m scan area, where the yellow regions represent the inverted domains. (d) Statistical analysis of the period $\Lambda$, inverted-domain width $W_{\rm R}$, and duty cycle $D$ of PPLN extracted from the PFM phase image in (c). The data points represent the mean values for 23 consecutive periods along the $x$ direction, while the shaded bands indicate the corresponding standard deviations, both calculated from row-by-row measurements along the $y$ direction. (e) 2D FFT of the PFM phase image in (c). The blue curve represents the corresponding Fourier coefficient profile extracted along $G_x$ at $G_y=0$.}
\end{figure*}

\begin{figure*}[htbp]
	\centering
	\includegraphics[width=15cm]{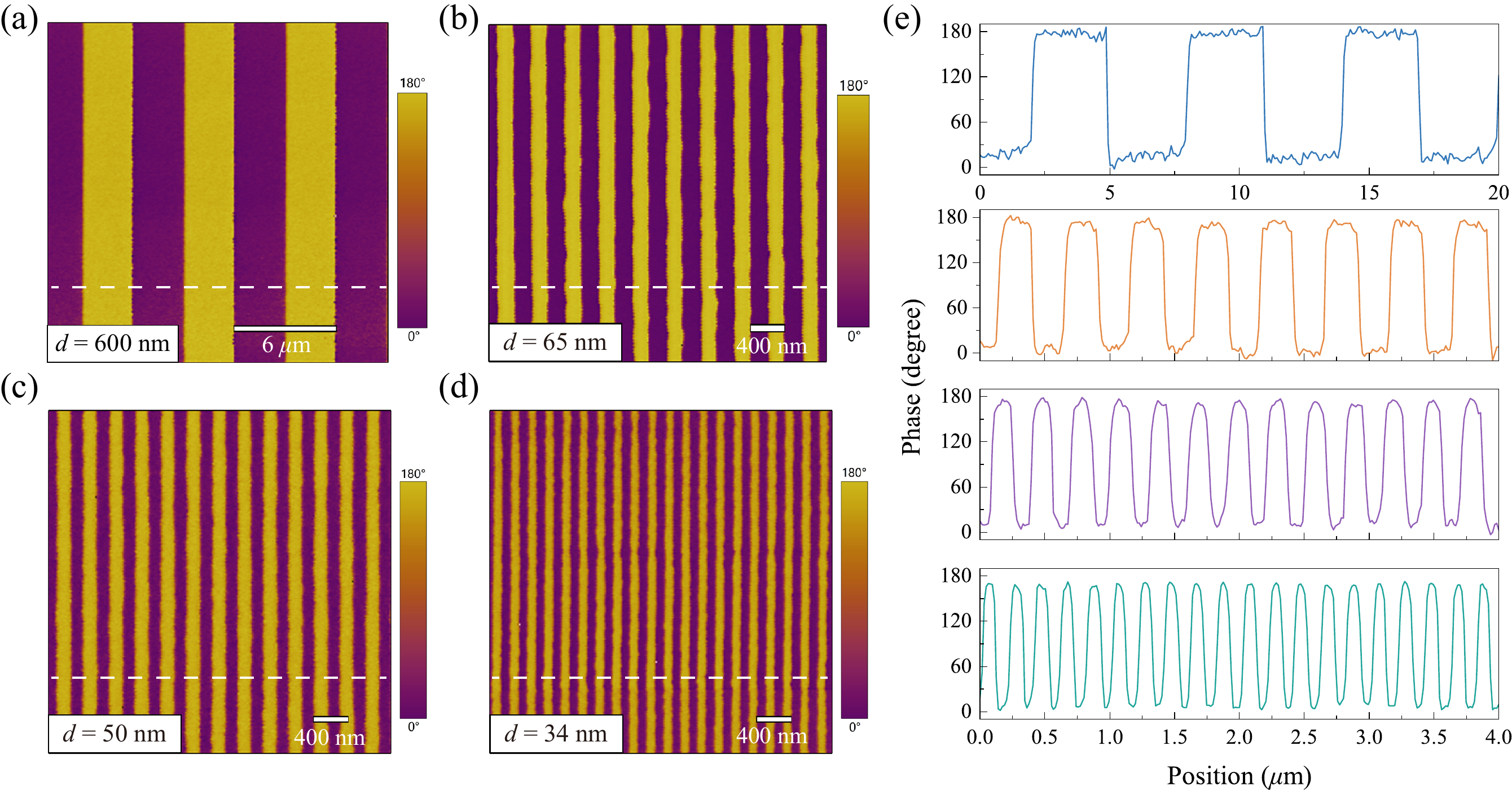}%
	\caption{\label{Fig5} Fabrication and characterization of PPLN structures with different periods. (a),(b) PFM phase images of PPLN structures with periods of 6 $\mu$m and 400 nm, respectively, fabricated in LNOSs using the voltage-ramping procedure described above. (c),(d) PFM phase images of PPLN structures with periods of 300 and 200 nm, respectively, fabricated using rectangular voltage pulses of 19 V (100 ms, three pulses) and 17 V (100 ms, one pulse), respectively. (e) PFM phase distribution profiles extracted along the white dashed lines in (a-d), arranged from top to bottom in the corresponding order. The LN film thickness $d$ is indicated in the bottom-left corner of each panel in (a-d), and the yellow regions correspond to the inverted domains.}
\end{figure*}

We further investigated the thickness-dependent polarization switching behavior by examining the coercive field $E_{\rm{c}}$, as shown in Fig.~\ref{Fig3}(a). The voltage corresponding to the peak in the poling current density was defined as the coercive voltage, and $E_{\rm{c}}$ was estimated by dividing the coercive voltage by the film thickness $d$. As the film thickness $d$ decreases, $E_{\rm{c}}$ increases markedly. To further reveal the scaling relationship, the results are replotted on a logarithmic scale in the inset of Fig.~\ref{Fig3}(a). A linear fit yields a slope of $-0.67$ over the film thickness range from 32 nm to 1170 nm, indicating that $E_{\rm{c}}$ follows the Kay-Dunn scaling law ($E_{\rm{c}} \propto d^{-2/3}$)~\cite{kay1962thickness}. Similar behaviors have also been reported in other ferroelectric materials~\cite{PhysRevLett.98.217602,41}.

We also examined the dependence of the inverted-domain broadening $\Delta W_{\rm{R}}$ on the film thickness $d$, with the poling electrode linewidth $W_{\rm E}$ fixed at 200 nm. Here, the inverted-domain broadening $\Delta W_{\rm{R}}$ is defined as the lateral size difference between the inverted domain and the electrode linewidth $W_{\rm E}$.  The values of $\Delta W_{\rm{R}}$ were extracted from PFM images, and  the results are summarized in Fig.~\ref{Fig3}(b), where each data point represents the averaged value of measurements   acquired from PFM images of five different poled regions at the same film thickness, and the error bars indicate the corresponding overall standard deviations. A general trend of increasing inverted-domain broadening $\Delta W_{\rm{R}}$ with increasing film thickness $d$ was observed. This trend can be attributed to the lateral expansion of inverted domains that accompanies their forward growth through the film thickness during polarization reversal, leading to more pronounced domain broadening in thicker films. 

The dependence of $\Delta W_{\rm{R}}$ on the poling electrode linewidth $W_{\rm{E}}$ was then investigated for different film thicknesses, and representative results for film thicknesses of 105, 377, 577, 729, 995, and 1170 nm are summarized in Fig.~\ref{Fig3}(c), and again the error bars represent the overall standard deviations obtained from five PFM images for each case. One sees that, for a given film thickness $d$, $\Delta W_{\rm{R}}$ remains approximately constant across different electrode linewidths, suggesting that the final inverted-domain width is primarily determined by the electrode linewidth and the associated lateral domain broadening. Similar behavior was observed for LNOSs with other film thicknesses. The corresponding duty cycles are labeled within each bar, illustrating the tunability of the duty cycle through electrode linewidth adjustment. Note that in all cases in Fig.~\ref{Fig3}, the period of the PPLN structures is 800 nm. These findings provide useful guidance for poling electrode design and suggest that lateral domain broadening can be pre-compensated by appropriately selecting the electrode linewidth to achieve the desired domain widths and duty cycles.

To further demonstrate the feasibility of fabricating extended PPLN structures with submicrometer periodicity, we fabricated a 3-mm-long PPLN structure with a period of 800 nm in the LNOS, where the LN film thickness $d$ was approximately 400 nm. The periodic Cr poling electrodes, with a period of 800 nm and a longitudinal length of 20 $\mu$m, were connected to a 40 $\mu$m $\times$ 40 $\mu$m contact pad through a 3-mm-long, 2-$\mu$m-wide Cr stripe. Based on the thickness-dependent domain broadening characterized above, the electrode linewidth was set to approximately 215 nm to pre-compensate for the lateral domain broadening during polarization reversal, aiming to achieve a duty cycle of approximately 50\% in the resulting PPLN structure. After poling, the Cr electrodes were removed, and the sample was cleaned as described above. Figure~\ref{Fig4}(a) shows an overview SEM image of the 3-mm-long PPLN structure, while Fig.~\ref{Fig4}(b) presents an enlarged SEM image of the region marked by the red box in Fig.~\ref{Fig4}(a), revealing well-defined periodic domain patterns. To further characterize the domain structure, PFM phase imaging was performed over a 20 $\mu$m $\times$ 20 $\mu$m area in the middle of the 3-mm-long structure, as shown in Fig.~\ref{Fig4}(c). The period $\Lambda$, inverted-domain width $W_{\rm R}$, and duty cycle $D$ of PPLN along the $x$ direction were extracted from the PFM phase image, which were statistically analyzed along the $y$ direction, as summarized in Fig.~\ref{Fig4}(d). The data points represent the mean values, while the shaded bands indicate the corresponding standard deviations calculated from row-by-row measurements along the $y$ direction. In this region, the inverted-domain width was approximately 400 nm, corresponding to a duty cycle close to 50\%. The two-dimensional fast Fourier transform (2D FFT) of the PFM phase image is presented in Fig.~\ref{Fig4}(e), with the blue curve showing the corresponding Fourier coefficient profile extracted along $G_x$ at $G_y=0$, where $G_{x,y}=2\pi/\Lambda_{x, y}$ is the spatial wavenumber of PPLN structure along the x or y direction. Two peaks are identified at spatial wavenumbers $G_x=\pm7.826$~$\mu$m$^{-1}$, corresponding to a period of 802.8 nm. These results further support the effectiveness of the lateral domain broadening pre-compensation strategy in achieving the desired duty cycle for a given LN film thickness.

The LNOS platform also provides a versatile platform for fabricating PPLN structures over a wide range of periods. As shown in the PFM phase image in Fig.~\ref{Fig5}(a), PPLN structure with a period of 6 $\mu$m was successfully fabricated in a 600-nm-thick LN film using a target poling voltage of 30 V and an electrode linewidth of 3 $\mu$m. The corresponding phase distribution profile extracted along the white dashed line is presented in the top panel of Fig.~\ref{Fig5}(e),  showing well-defined periodic domains with good structural fidelity. As discussed in Figs.~\ref{Fig3}(b) and \ref{Fig3}(c), lateral domain broadening becomes more pronounced with increasing film thickness. Therefore, reducing the film thickness and poling electrode linewidth is beneficial for fabricating PPLN structures with smaller periods. By reducing the film thickness to 65 nm and the electrode linewidth to 90 nm, PPLN structure with a period of 400 nm was successfully fabricated, as shown in Fig.~\ref{Fig5}(b). To achieve even smaller periods, rectangular voltage pulses were employed instead of the voltage ramping procedure described above. As shown in Figs.~\ref{Fig5}(c) and \ref{Fig5}(d), PPLN structures with periods of 300 and 200 nm were fabricated in the 50- and 34-nm-thick LN films using rectangular voltage pulses of 19 V (100 ms, three pulses) and 17 V (100 ms, one pulse), respectively. The corresponding phase distribution profiles of the 400-, 300-, and 200-nm-period PPLN structures are presented in the second, third, and bottom panels of Fig.~\ref{Fig5}(e), respectively.

These results demonstrate the capability of the LNOS platform to support the fabrication of PPLN structures with periods ranging from the micrometer to the submicrometer scale. In particular, the combination of reduced film thickness, narrow poling electrodes, and pulsed-voltage poling enables the fabrication of PPLN structures with periods down to 200 nm. To the best of our knowledge, this represents one of the smallest periods reported for PPLN structures fabricated by electric-field poling. Notably, PPLN structures with submicrometer periodicity retained their periodic domain patterns after several months, even for the structure with a period as small as 200 nm, as confirmed by subsequent PFM measurements, demonstrating their long-term domain stability.

\section{Conclusions}

In summary, we have demonstrated the fabrication of monocrystalline LN films with thicknesses ranging from micrometers down to tens of nanometers on low-resistivity silicon substrates through chemical-mechanical polishing. Using electric-field poling technique, high-quality PPLN structures with submicrometer periodicity were successfully fabricated with full-thickness polarization reversal and well-defined depth profiles. The thickness-dependent coercive field was found to be consistent with the Kay-Dunn scaling law, while the lateral domain broadening in general increased with increasing film thickness, but remained approximately constant across different electrode linewidths at a given film thickness. PPLN structures with periods ranging from 6 $\mu$m down to 200 nm were demonstrated, including a 3-mm-long structure with an 800-nm period. The observed domain broadening behavior provides guidance for poling electrode design, allowing the electrode linewidth to be adjusted in advance to pre-compensate for lateral domain expansion and therefore achieve the desired domain widths and duty cycles. These results offer a practical route toward high-quality and scalable domain engineering in LN films, with promising potential for extending quasi-phase-matching capabilities and enabling compact, highly integrated nonlinear photonic devices.

\begin{acknowledgments}
This work is supported by the National Key Research and Development Program of China (Grant No. 2022YFA1404604), the National Natural Science Foundation of China (NSFC) (Grant No. 12134007), and the Higher Education Discipline Innovation Project (Grant No. B23045).
\end{acknowledgments}
\medskip
\noindent\textbf{DATA AVAILABILITY}

Data may be obtained from the authors upon reasonable request.

%\appendix

%\section{Appendixes}

%If there is only one appendix, then the letter ``A'' should not appear. This is suppressed by using the star version of the appendix command (\verb+\appendix*+ in the place of \verb+\appendix+).

%\section{A little more on appendixes}

% The \nocite command causes all entries in a bibliography to be printed out
% whether or not they are actually referenced in the text. This is appropriate
% for the sample file to show the different styles of references, but authors
% most likely will not want to use it.
%

%\bibliography{nanoscalePPLN}% Produces the bibliography via BibTeX.

\end{document}